# On the Complexity of Dynamic Epistemic Logic [*]

Guillaume Aucher
University of Rennes 1 - INRIA
guillaume.aucher@irisa.fr

François Schwarzentruber
ENS Cachan - Brittany extension
francois.schwarzentruber@bretagne.ens-cachan.fr

## ABSTRACT

Although Dynamic Epistemic Logic (DEL) is an influential logical framework for representing and reasoning about information change, little is known about the computational complexity of its associated decision problems. In fact, we only know that for public announcement logic, a fragment of DEL, the satisfiability problem and the model-checking problem are respectively PSPACE-complete and in P. We contribute to fill this gap by proving that for the DEL language with event models, the model-checking problem is, surprisingly, PSPACE-complete. Also, we prove that the satisfiability problem is NEXPTIME-complete. In doing so, we provide a sound and complete tableau method deciding the satisfiability problem.

## Categories and Subject Descriptors

I.2.4 [**Knowledge representation formalisms and methods**]: Modal logic; F.1.3 [**Complexity measure and classes**]: Reducibility and completeness

## General Terms

Theory

## 1. INTRODUCTION

Research fields like distributed artificial intelligence, distributed computing and game theory all deal with groups of human or non-human agents which interact, exchange and receive information. The problems they address range from multi-agent planning and design of distributed protocols to strategic decision making in groups. In order to address appropriately and rigorously these problems, it is necessary to be able to provide formal means for representing and reasoning about such interactions and flows of information. The framework of Dynamic Epistemic Logic (DEL for short) is very well suited to this aim. Indeed, it is a logical framework where one can represent and reason about beliefs and knowledge change of multiple agents, and more generally about information change.

The theoretical work of the above mentioned research fields has already been applied to various practical problems stemming from telecommunication networks, World Wide Web, peer to peer networks, aircraft control systems, and so on... In general, in all applied contexts, the investigation of the algorithmic aspects of the formalisms employed plays an important role in determining whether and to what extent they can be applied. For this reason, the algorithmic aspects of DEL need to be studied.

To this aim, a preliminary step consists in studying the computational properties of its main associated decision problems, namely the model checking problem and the satisfiability problem. Indeed, it will indirectly provide algorithmic methods to solve these decision problems and give us a hint of whether and to what extent our methods can be applied. However, surprisingly little is known about the computational complexity of these problems. We only know that for public announcement logic, a fragment of DEL [Plaza, 1989], the model checking problem is in P and the satisfiability problem is PSPACE-complete. Here, we aim to fill this gap for the full language of DEL with event models.

DEL is built on top of epistemic logic. An epistemic model represents how a given set of agents perceive the actual world in terms of beliefs and knowledge about this world and about the other agents' beliefs. The insight of the DEL approach is that one can describe how an event is perceived by agents in a very similar way: an agent's perception of an event can also be described in terms of beliefs and knowledge. For example, at the battle of Waterloo, when marshal Blücher received the message of the duke of Wellington inviting him to join the attack at dawn against Napoleon, Wellington did not *know* at that very moment that Blücher was receiving his message, and Blücher *knew* it. This is a typical example of announcement which is not public. This led Baltag, Moss and Solecki to introduce the notion of *event model* [Baltag et al., 1998]. The definition of an event model, denoted $(\mathcal{M}', w')$, is very similar to the definition of an epistemic model. They also introduced a *product update*, which defines a new epistemic model representing the situation after the event. Then, they extended the epistemic language with dynamic operators $[\mathcal{M}', w']\varphi$ standing for '$\varphi$ holds after the occurrence of the event represented by $(\mathcal{M}', w')$'.

Using the so-called reduction axioms, it turns out that any formula with dynamic operator(s) can be translated to an equivalent epistemic formula without dynamic operator. As a first approximation, we could be tempted to use these reduction axioms to reduce both the model checking problem and the satisfiability problem of DEL to the model checking problem and the satisfiability problem of epistemic logic, because optimal algorithmic methods already exist for these related problems. However, the reduction algorithm induced by the reduction axioms is exponential in the size of the input formula. Therefore, for the satisfiability problem, we only obtain an algorithm which is in EXPSPACE (because the satisfiability problem of epistemic logic is PSPACE-complete), and for the model checking problem, we only obtain an algorithm which is in EXPTIME (because the model checking problem of epistemic logic is in P). These algorithms are not optimal because, as we shall see, there exists an algorithm solving the satisfiability problem which is in NEXPTIME⊆ EXPSPACE and also an algorithm solving the model checking problem which is in PSPACE⊆ EXPTIME. Our algorithm for solving the satisfiability problem is based on a sound and complete tableau method which does not resort to the reduction axioms.

---

[*] An extended version of this article with full proofs can be found at the following url: http://hal.inria.fr/docs/00/75/95/44/PDF/RR-8164.pdf

The paper is organized as follows. In Section 2, we recall the core of the DEL framework and the different variants of languages with event models which have been introduced in the literature. In Section 3, we prove that the model checking problem of DEL is PSPACE-complete, and in Section 4 we prove that the satisfiability problem is NEXPTIME-complete. In Section 5, we discuss related works and whether our results still hold when we extend the expressiveness of the language with common belief and 'star' iteration operators. We conclude in Section 6.

## 2. DYNAMIC EPISTEMIC LOGIC

Following the methodology of DEL, we split the exposition of the DEL logical framework into three subsections. In Section 2.1, we recall the syntax and semantics of the epistemic language. In Section 2.2, we define event models, and in Section 2.3, we define the product update. In Section 2.4, we recall the different languages that have been introduced in the DEL literature and we introduce our language $\mathcal{L}_{DEL}$.

### 2.1 Epistemic language

In the rest of the paper, $ATM$ is a countable set of atomic propositions and $AGT$ is a finite set of agents.

A (pointed) epistemic model $(\mathcal{M}, w)$ represents how the actual world represented by $w$ is perceived by the agents. Intuitively, in this definition, $vR_a u$ means that in world $v$ agent $a$ considers that world $u$ might be the actual world.

DEFINITION 1 (EPISTEMIC MODEL).
An *epistemic model* is a tuple $\mathcal{M} = (W, R, V)$ where $W$ is a non-empty set of possible worlds, $R$ maps each agent $a \in AGT$ to a relation $R_a \subseteq W \times W$ and $V : ATM \to 2^W$ is a function called a valuation. We abusively write $w \in \mathcal{M}$ for $w \in W$ and we say that $(\mathcal{M}, w)$ is a *pointed epistemic model*. We also write $v \in R_a(w)$ for $wR_a v$.

Then, we define the following epistemic language $\mathcal{L}_{EL}$. It can be used to state properties of epistemic models:

DEFINITION 2 (EPISTEMIC LANGUAGE).
The *language* $\mathcal{L}_{EL}$ of epistemic logic is defined as follows:

$$\mathcal{L}_{EL} : \varphi ::= p \mid \neg \varphi \mid (\varphi \wedge \varphi) \mid B_a \varphi$$

where $p$ ranges over $ATM$ and $a$ ranges over $AGT$. A formula of $\mathcal{L}_{EL}$ is called an *epistemic formula*. The formula $\bot$

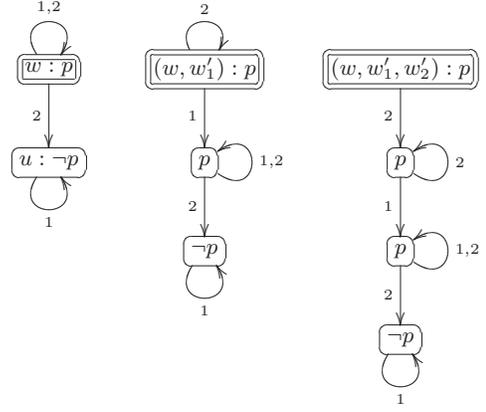

Figure 1: Pointed epistemic models $(\mathcal{M}, w)$ (*left*), $((\mathcal{M} \otimes \mathcal{M}'_1), (w, w'_1))$ (*center*) and $(\mathcal{M} \otimes \mathcal{M}'_1 \otimes \mathcal{M}'_2, (w, w'_1, w'_2))$ (*right*)

is an abbreviation for $p \wedge \neg p$, and $\top$ is an abbreviation for $\neg \bot$. The formula $\langle B_a \rangle \varphi$ is an abbreviation of $\neg B_a \neg \varphi$. The *size of a formula* $\varphi \in \mathcal{L}_{EL}$ is defined by induction as follows: $|p| = 1$; $|\neg \varphi| = 1 + |\varphi|$; $|\varphi \wedge \psi| = 1 + |\varphi| + |\psi|$; $|B_a \varphi| = 1 + |\varphi|$.

Intuitively, the formula $B_a \varphi$ reads as 'agent $a$ believes that $\varphi$ holds in the current situation'.

DEFINITION 3 (TRUTH CONDITIONS).
Given an epistemic model $\mathcal{M} = (W, R, V)$ and a formula $\varphi \in \mathcal{L}_{EL}$, we define inductively the satisfaction relation $\models \subseteq W \times \mathcal{L}_{EL}$ as follows: for all $w \in W$,

$\mathcal{M}, w \models p$ iff $w \in V(p)$
$\mathcal{M}, w \models \varphi \wedge \psi$ iff $\mathcal{M}, w \models \varphi$ and $\mathcal{M}, w \models \psi$
$\mathcal{M}, w \models \neg \varphi$ iff not $\mathcal{M}, w \models \varphi$
$\mathcal{M}, w \models B_a \varphi$ iff for all $v \in R_a(w)$, we have $\mathcal{M}, v \models \varphi$

We write $\mathcal{M} \models \varphi$ when for all $w \in \mathcal{M}$, it holds that $\mathcal{M}, w \models \varphi$. Also, we write $\models \varphi$, and we say that $\varphi$ is *valid*, when for all epistemic model $\mathcal{M}$, it holds that $\mathcal{M} \models \varphi$. Dually, we say that $\varphi$ is *satisfiable* when $\neg \varphi$ is not valid.

EXAMPLE 1. *Our running example is inspired by the coordinated attack problem from the distributed systems folklore [Fagin et al., 1995]. Our set of atomic propositions is $ATM = \{p\}$ and our set of agents is $AGT = \{1, 2\}$. Agent 1 is the duke of Wellington and agent 2 is marshal Blücher; $p$ stands for 'Wellington wants to attack at dawn'. The initial situation is represented in Figure 1 by the pointed epistemic model $(\mathcal{M}, w) = (\{w, u\}, R_1 = \{(w, w), (u, u)\}, R_2 = \{(w, w), (w, u)\}, V(p) = \{w\})$. In this pointed epistemic model, it holds that $\mathcal{M}, w \models p \wedge B_1 p$: Wellington 'knows' that he wants to attack at dawn. It also holds that $\mathcal{M}, w \models \neg B_2 p$: Blücher does not 'know' that Wellington wants to attack at dawn; and $\mathcal{M}, w \models B_1 \neg B_2 p$: Wellington 'knows' that Blücher does not 'know' that he wants to attack at dawn.*

### 2.2 Event model

A (pointed) event model $(\mathcal{M}', w')$ represents how the actual event represented by $w'$ is perceived by the agents. Intuitively, in this definition, $u' R'_a v'$ means that while the possible event represented by $u'$ is occurring, agent $a$ considers possible that the event represented by $v'$ is in fact occurring.

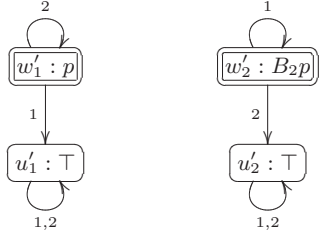

**Figure 2:** Pointed event models $(\mathcal{M}'_1, w'_1)$ (*left*) and $(\mathcal{M}'_2, w'_2)$ (*right*)

### 2.3 Product update

DEFINITION 4 (EVENT MODEL).
An *event model* is a tuple $\mathcal{M}' = (W', R', Pre)$ where $W'$ is a non-empty and finite set of possible events, $R'$ maps each agent $a \in AGT$ to a relation $R'_a \subseteq W' \times W'$ and $Pre : W' \to \mathcal{L}_{EL}$ is a function that maps each event to a precondition expressed in the epistemic language.

We abusively write $w' \in \mathcal{M}'$ for $w' \in W'$ and we say that $(\mathcal{M}', w')$ is a *pointed event model*. The *size of an event model* $\mathcal{M}' = (W', R', Pre)$ is noted $|\mathcal{M}'|$ and is defined as follows: $card(W') + \sum_{a \in AGT} card(R'_a) + \sum_{w' \in W'} |Pre(w')|$.

EXAMPLE 2. *In Figure 2 are represented two pointed event models. The first, $(\mathcal{M}_1, w'_1) = (\{w'_1, u'_1\}, R_1 = \{(w'_1, u'_1), (u'_1, u'_1)\}, R_2 = \{(w'_1, w'_1), (u'_1, u'_1)\}, Pre, w'_1)$ where $Pre(w'_1) = p$ and $Pre(u'_1) = \top$, represents the event whereby Blücher receives the message of Wellington that he wants to attack at dawn. When this happens, Wellington believes that nothing happens and believes that this is even common knowledge. The second, $(\mathcal{M}_2, w'_2) = (\{w'_2, u'_2\}, R_1 = \{(w'_2, w'_2), (u'_2, u'_2)\}, R_2 = \{(w'_2, u'_2), (u'_2, u'_2)\}, Pre, w'_2)$, where $Pre(w'_2) = B_2 p$ and $Pre(u'_2) = \top$, represents the event whereby Wellington receives the message of Blücher telling him that he 'knows' that Wellington wants to attack at dawn.*

### 2.3 Product update

The following product update yields a new pointed epistemic model $\mathcal{M} \otimes \mathcal{M}', (w, w')$ representing how the new situation which was previously represented by $(\mathcal{M}, w)$ is perceived by the agents after the occurrence of the event represented by $(\mathcal{M}', w')$.

DEFINITION 5 (PRODUCT UPDATE).
Let $\mathcal{M} = (W, R, V)$ be an epistemic model and let $\mathcal{M}' = (W', R', Pre)$ be an event model. The *product update of $\mathcal{M}$ by $\mathcal{M}'$* is the epistemic model $\mathcal{M} \otimes \mathcal{M}' = (W'', R'', V'')$ defined as follows ($p$ and $a$ range over $ATM$ and $AGT$ respectively):

$$W'' = \{(w, w') \in W \times W' \mid \mathcal{M}, w \models Pre(u')\}$$
$$R''_a = \{\langle (w, w'), (v, v') \rangle \in W'' \times W'' \mid wR_a v \text{ and } w'R'_a v'\}$$
$$V''(p) = \{(w, w') \in W'' \mid w \in V(p)\}$$

Given a pointed epistemic model $(\mathcal{M}, w)$, and a pointed event model $(\mathcal{M}', w')$, we say that $(\mathcal{M}', w')$ is *executable* in $(\mathcal{M}, w)$ when $\mathcal{M}, w \models Pre(w')$. If $\mathcal{M}$ is an epistemic model and $\mathcal{M}'_1, \ldots, \mathcal{M}'_n$ are event models, we abusively write $\mathcal{M} \otimes \mathcal{M}'_1 \otimes \cdots \otimes \mathcal{M}'_n$ for $(\ldots((\mathcal{M} \otimes \mathcal{M}'_1) \otimes \mathcal{M}'_2) \otimes \ldots) \otimes \mathcal{M}'_n$ and $(w, w'_1, \ldots, w'_n)$ for $(\ldots((w, w'_1), w'_2), \ldots), w'_n)$.

EXAMPLE 3. *The pointed epistemic models $((\mathcal{M} \otimes \mathcal{M}'_1), (w, w'_1))$ and $(\mathcal{M} \otimes \mathcal{M}'_1 \otimes \mathcal{M}'_2, (w, w'_1, w'_2))$ are represented in Figure 1. After Blücher receives the message of Wellington, Blücher 'knows' that Wellington wants to attack at dawn, but Wellington does not 'know' that Blücher 'knows' it: $\mathcal{M} \otimes \mathcal{M}'_1, (w, w'_1) \models p \wedge B_2 p \wedge \neg B_1 B_2 p$. Likewise, after Wellington receives the message of Blücher telling him that he 'knows' that he wants to attack at dawn ($B_2 p$), Wellington 'knows' that Blücher 'knows' that he wants to attack at dawn, but Blücher does not 'know' that Wellington 'knows' it: $\mathcal{M} \otimes \mathcal{M}'_1 \otimes \mathcal{M}'_2, (w, w'_1, w'_2) \models p \wedge B_2 p \wedge B_1 B_2 p \wedge \neg B_2 B_1 B_2 p$. Hence, in particular, $\mathcal{M}, w \models \neg [\mathcal{M}'_1, w'_1][\mathcal{M}'_2, w'_2] B_2 B_1 B_2 p$.*

### 2.4 Languages of DEL

In [Baltag et al., 1998], the language is defined as follows:

$$\varphi ::= p \mid \neg \varphi \mid (\varphi \wedge \varphi) \mid B_a \varphi \mid [\mathcal{M}', w'] \varphi$$

where $p$ ranges over $ATM$, $a$ over $AGT$ and $(\mathcal{M}', w')$ is any pointed and finite event model. The formula $\langle \mathcal{M}', w' \rangle \varphi$ is an abbreviation for $\neg [\mathcal{M}', w'] \neg \varphi$.

Intuitively, $[\mathcal{M}', w'] \varphi$ reads as '$\varphi$ will hold after the occurrence of the event represented by $(\mathcal{M}', w')$' and $\langle \mathcal{M}', w' \rangle \varphi$ reads as 'the event represented by $(\mathcal{M}', w')$ is executable in the current situation and $\varphi$ will hold after its execution'.

However, note that in this definition, preconditions of event models are necessarily epistemic formulas. In [Baltag and Moss, 2004], another language is introduced which can deal with event models whose preconditions may involve formulas with event models. This language relies on the notion of *event signature* and the epistemic language is extended with a modality $[\Sigma, \varphi_1, \ldots, \varphi_n] \varphi$, where $\Sigma$ is an event signature. The language of [Baltag and Moss, 2004] also includes PDL-like program constructions such as sequential composition, union and 'star' operation of event models (see Section 5 for a definition of these program constructions).

In [van Ditmarsch et al., 2007], preconditions can also be formulas involving event models, but only union of programs is allowed. It is therefore a fragment of the language of [Baltag and Moss, 2004] since it does not include sequential composition nor the 'star' operation. This will be our language in this paper.

DEFINITION 6 ([VAN DITMARSCH ET AL., 2007]).
The language $\mathcal{L}_{DEL}$ is the union of the *formulas* $\varphi \in \mathcal{L}^{stat}_\otimes$ and the *events* (or *epistemic events*) $\pi \in \mathcal{L}^{dyn}_\otimes$ defined by the following rule:

$$\mathcal{L}^{stat}_\otimes : \varphi ::= p \mid \neg \varphi \mid (\varphi \wedge \varphi) \mid B_a \varphi \mid [\pi] \varphi$$
$$\mathcal{L}^{dyn}_\otimes : \pi ::= \mathcal{M}', w' \mid (\pi \cup \pi)$$

where $p$ ranges over $ATM$, $a$ over $AGT$ and $(\mathcal{M}', w')$ is any pointed and finite event model such that for all $w' \in \mathcal{M}'$, $Pre(w')$ is a formula of $\mathcal{L}^{stat}_\otimes$ that has already been constructed in a previous stage of the inductively defined hierarchy.

The size of $\varphi \in \mathcal{L}_{DEL}$ is defined as for the epistemic language together with the induction case $|[\pi] \varphi| = 1 + |\pi| + |\varphi|$ where $|\mathcal{M}', w'| = |\mathcal{M}'|$, and $|\pi \cup \gamma| = 1 + |\pi| + |\gamma|$.

DEFINITION 7 (TRUTH CONDITIONS).
Given an epistemic model $\mathcal{M} = (W, R, V)$ and a formula $\varphi \in \mathcal{L}_{DEL}$, we define inductively the satisfaction relation

$\models \; \subseteq W \times \mathcal{L}_{DEL}$ as follows:

$\mathcal{M}, w \models [\mathcal{M}', w']\varphi$ iff $\mathcal{M}, w \models Pre(w')$ implies
$\mathcal{M} \otimes \mathcal{M}', (w, w') \models \varphi$
$\mathcal{M}, w \models [\pi \cup \gamma]\varphi$ iff $\mathcal{M}, w \models [\pi]\varphi$ and $\mathcal{M}, w \models [\gamma]\varphi$.

The other induction steps are identical to the induction steps of Definition 3.

The results in this paper are the same whether or not the formulas of the preconditions involve event models. However, the result of NEXPTIME-completeness of the satisfiability problem of Section 4 holds only if we consider union of event models as a program construction in the language.

## 3. MODEL CHECKING PROBLEM

The *model checking problem of* $\mathcal{L}_{DEL}$ is defined as follows:

**Input**: a pointed epistemic model $(\mathcal{M}, w)$ and a formula $\varphi \in \mathcal{L}_{DEL}$;

**Output**: yes iff $\mathcal{M}, w \models \varphi$.

Whereas the model checking problem with an epistemic formula of $\mathcal{L}_{EL}$ is in P, model checking with a formula of $\mathcal{L}_{DEL}$ is surprisingly PSPACE-complete. This shows that the addition of dynamic modalities with event models to $\mathcal{L}_{EL}$ increases tremendously the computational complexity of the model checking problem.

### 3.1 Upper bound

In Figure 3 is defined a deterministic algorithm M-Check($w \; \mathcal{M}'_1, w'_1; \ldots; \mathcal{M}'_i, w'_i \; , \varphi$) that checks whether we have $\mathcal{M} \otimes \mathcal{M}'_1 \otimes \ldots \mathcal{M}'_i, (w, w'_1, \ldots, w'_i) \models \varphi$, where $(\mathcal{M}, w)$ is a pointed epistemic model and for all $j \in \{1, \ldots, i\}$, $(\mathcal{M}'_j, w'_j)$ is a pointed event model. The precondition of a call to the function M-Check($w \; \mathcal{M}'_1, w'_1; \ldots; \mathcal{M}'_i, w'_i \; , \varphi$) is that $(w, w'_1, \ldots, w'_i) \in \mathcal{M} \otimes \mathcal{M}'_1 \otimes \ldots \mathcal{M}'_i$, that is, the sequence $(\mathcal{M}'_1, w'_1) \ldots (\mathcal{M}'_i, w'_i)$ is executable in $(\mathcal{M}, w)$. In order to check whether $\mathcal{M}, w \models \varphi$, we just call M-Check($w, \varphi$).

THEOREM 1. *The model checking problem of $\mathcal{L}_{DEL}$ is in PSPACE.*

PROOF SKETCH. Termination and correction of the algorithm M-Check are easily proved over the size of the input defined by $|\mathcal{M}| + \sum_{k=1}^{i} |\mathcal{M}'_k| + |\varphi|$. As for complexity, the algorithm requires a polynomial amount of space in the size of the input. Indeed, as the size of the input is strictly decreasing at each recursive call, the number of recursive calls in the call stack is linear in the size of the input. Then, each of the current call requires a polynomial amount of space in the size of the input for storing the value of local variables: the most consuming case is $B_a\psi$ where we have to save all the current values of $u, u_1, \ldots, u_i$ in the loop **for**. □

### 3.2 Lower bound

We prove that the algorithm of the previous section is optimal. To do so, we provide a polynomial reduction of the *quantified Boolean formula satisfiability problem*, known to be PSPACE-complete [Papadimitriou, 1995, p. 455] to the model-checking problem of $\mathcal{L}_{DEL}$.

```
function M-Check(w  M'_1,w'_1;...;M'_i,w'_i  φ)
  match (φ)
    case p:
      return w ∈ V(p);
    case ¬ψ:
      return not M-Check(w  M'_1,w'_1;...;M'_i,w'_i  ψ);
    case ψ_1 ∧ ψ_2:
      return (M-Check(w  M'_1,w'_1;...;M'_i,w'_i  ψ_1) and
        M-Check(w  M'_1,w'_1;...;M'_i,w'_i  ψ_2));
    case B_aψ:
      for u ∈ R_a(w)
      for u'_1 ∈ R'_a(w'_1)
      if M-Check(u, Pre(u'_1))
        .
        .
        .
      for u'_i ∈ R'_a(w'_i)
      if M-Check(u  M'_1,u'_1;...;M'_{i-1},u'_{i-1}  Pre(u'_i))
        if not M-Check(u  M'_1,u'_1;...;M'_i,u'_i  ψ);
          return false ;
        endIf
      endIf endFor ...endIf endFor endFor
      return true ;
    case [M',w']ψ:
      if M-Check(w  M'_1,w'_1;...;M'_i,w'_i  Pre(w'))
        return M-Check(w  M'_1,w'_1;...;M'_i,w'_i;M',w'  ψ);
      endIf
      return true ;
    case [π ∪ γ]ψ:
      return (M-Check(w  M'_1,w'_1;...;M'_i,w'_i  [π]ψ) and
        M-Check(w  M'_1,w'_1;...;M'_i,w'_i  [γ]ψ));
  endMatch
endFunction
```

**Figure 3: PSPACE algorithm for model checking**

THEOREM 2. *The model checking problem of $\mathcal{L}_{DEL}$ is PSPACE-hard.*

PROOF. Without loss of generality, we only consider in this proof quantified Boolean formulas of the form $\forall p_1 \exists p_2 \forall p_3 \ldots \forall p_{2k-1} \exists p_{2k} \psi(p_1, \ldots p_{2k})$, where $\psi(p_1, \ldots, p_{2k})$ is a Boolean formula over the atomic propositions $p_1, \ldots, p_{2k}$. The formula $\forall p_1 \exists p_2 \forall p_3 \ldots \forall p_{2k-1} \exists p_{2k} \psi(p_1, \ldots p_{2k})$ is *satisfiable* iff for both truth values of the atomic proposition $p_1$ there is a truth value for the atomic proposition $p_2$ such that for both truth values of the atomic proposition $p_3$, and so on up to $p_{2k}$, the formula $\psi(p_1, \ldots p_{2k})$ is true in the overall truth assignment.

We can restrict ourselves to $\mathcal{L}_{DEL}$ where there is only one agent $a$. The *quantified Boolean formula satisfiability problem* is defined as follows:

**Input**: a natural number $k$ and a quantified Boolean formula $\varphi \triangleq \forall p_1 \exists p_2 \forall p_3 \ldots \forall p_{2k-1} \exists p_{2k} \psi(p_1, \ldots, p_{2k})$;

**Output**: yes iff $\varphi$ is satisfiable.

Let $\varphi = \forall p_1 \exists p_2 \forall p_3 \ldots \forall p_{2k-1} \exists p_{2k} \psi(p_1, \ldots p_{2k})$ be a quantified Boolean formula. We define a pointed epistemic model $(\mathcal{M}, w^0)$, $2k$ pointed event models $(\mathcal{M}'_1, w'^0_1), \ldots, (\mathcal{M}'_{2k}, w'^0_{2k})$, a pointed event model $\mathcal{M}'_\circlearrowleft, w'^0_\circlearrowleft$ and an epistemic formula $\psi'$ that are computable in polynomial time in the size of $\varphi$ such that:

$\varphi$ is satisfiable in quantified Boolean logic
iff
$\mathcal{M}, w^0 \models [\mathcal{M}'_1, w'^0_1 \cup \mathcal{M}'_\circlearrowleft, w'^0_\circlearrowleft]\langle \mathcal{M}'_2, w'^0_2 \cup \mathcal{M}'_\circlearrowleft, w'^0_\circlearrowleft \rangle \ldots$
$[\mathcal{M}'_{2k-1}, w'^0_{2k-1} \cup \mathcal{M}'_\circlearrowleft, w'^0_\circlearrowleft]\langle \mathcal{M}'_{2k}, w'^0_{2k} \cup \mathcal{M}'_\circlearrowleft, w'^0_\circlearrowleft \rangle \psi'$.

The corresponding instance of the model checking problem of $\mathcal{L}_{DEL}$ is computable in polynomial time in the size of $\varphi$. Now, let us describe $\mathcal{M}, w_0$, the event models $\mathcal{M}'_1, w'^0_1, \ldots, \mathcal{M}'_{2k}, w'^0_{2k}$, the event model $\mathcal{M}'_\circlearrowleft, w'^0_\circlearrowleft$ and $\psi'$.

- $\mathcal{M} = (W, R, V)$ is defined by:
  - $W = \{w^0, w^1, \ldots, w^{2k+1}\}$;
  - $R_a = \{(w^j, w^{j+1}) \mid j \in \{0, \ldots, 2k\}\}$;
  - and $V(p) = \emptyset$ for all $p \in ATM$

- For all $i \in \{1, \ldots, 2k\}$, $\mathcal{M}'_i = (W'_i, R'_i, Pre_i)$ is defined by:
  - $W'_i = \{w'^0_i, w'^1_i, \ldots, w'^i_i, w'^\circlearrowleft_i\}$
  - $R'_{i_a} = \{(w'^j_i, w'^{j+1}_i) \mid j \in \{0, \ldots, i-1\}\} \cup \{(w'^0_i, w'^\circlearrowleft_i), (w'^\circlearrowleft_i, w'^\circlearrowleft_i)\}$
  - and $Pre_i(u') = \top$ for all $u' \in W'_i$

- $\mathcal{M}'_\circlearrowleft, w'^0_\circlearrowleft = (W'_\circlearrowleft, R'_\circlearrowleft, Pre_\circlearrowleft)$ is defined by:
  - $W'_\circlearrowleft = \{w'^0_\circlearrowleft\}$
  - $R'_{\circlearrowleft_a} = \{(w'^0_\circlearrowleft, w'^0_\circlearrowleft)\}$
  - $Pre_\circlearrowleft(w'^0_\circlearrowleft) = \top$

- $\psi' = \psi(p_1 \leftarrow \langle B_a \rangle B_a \bot, \ldots, p_{2k} \leftarrow (\langle B_a \rangle)^{2k} B_a \bot)$, that is, $\psi'$ is the formula $\psi$ where all $p_i$ occurrences are substituted by $(\langle B_a \rangle)^i B_a \bot$.[1]

The semantics is simulated in the following way. The proposition $p_i$ is interpreted as the presence of a chain of length exactly $i$ from the root of a given epistemic model. That is why in $\psi'$, the proposition $p_i$ is substituted by $(\langle B_a \rangle)^i B_a \bot$, which is true in the root of the final epistemic model iff there exists a chain of length $i$ in that model.

Note that updating an epistemic model where there is a chain of length $2k+1$ by $\mathcal{M}'_i, w'^0_i$ where $i \in \{1, \ldots, 2k\}$:

- preserves the presence or absence of any chain of length $j \neq i$; in particular, it always preserves the presence of the chain of length $2k+1$;
- adds a chain of length $i$, that is $p_i$ becomes true;

Note also that updating an epistemic model where there is a chain of length $2k+1$ by $\mathcal{M}'_\circlearrowleft, w'^0_\circlearrowleft$ preserves the presence or absence of any chain. So, it will keep $p_i$ false if it was already false and it will keep any $p_i$ true if it was already true. In other words, the $\mathcal{M}'_\circlearrowleft, w'^0_\circlearrowleft$ is a neutral element for the product update.

The crucial invariant property $(Inv)$ of an epistemic model is the existence of a chain of length $2k+1$ in any update of $\mathcal{M}, w^0$ by any sequence of $\mathcal{M}'_\circlearrowleft, w'^0_\circlearrowleft$ and $\mathcal{M}'_i, w'^0_i$.

The behavior of $\forall p_i$ in quantified Boolean logic consists in a universal choice of a truth value for $p_i$. It is translated by the update operator $[\mathcal{M}'_i, w'^0_i \cup \mathcal{M}'_\circlearrowleft, w'^0_\circlearrowleft]$ whose semantics is to choose *universally* the update of the epistemic model by $\mathcal{M}'_i, w'^0_i$, that will give a new updated epistemic model with a chain of length $i$, that is $p_i$ is true, or by $\mathcal{M}'_\circlearrowleft, w'^0_\circlearrowleft$ that will let the new updated epistemic model without a chain of length $i$, that is $p_i$ is false.

---

[1] The formula $(\langle B_a \rangle)^i \varphi$ is an abbreviation of $\underbrace{\langle B_a \rangle \ldots \langle B_a \rangle}_{i \ times} \varphi$.

The behavior of $\exists p_i$ in quantified Boolean logic consists in an existential choice of a truth value for $p_i$. It is translated by the update operator $\langle \mathcal{M}'_i, w'^0_i \cup \mathcal{M}'_\circlearrowleft, w'^0_\circlearrowleft \rangle$ whose semantics is to choose *existentially* the update of the epistemic model by $\mathcal{M}'_i, w'^0_i$, that will give a new updated epistemic model with a chain of length $i$, that is $p_i$ is true, or by $\mathcal{M}'_\circlearrowleft, w'^0_\circlearrowleft$, that will let the new updated epistemic model without a chain of length $i$, that is $p_i$ is false. □

REMARK 1. *Note that the reduction used to prove that the model checking problem of $\mathcal{L}_{DEL}$ is PSPACE-hard uses only the precondition $\top$.*

## 4. SATISFIABILITY PROBLEM

The *satisfiability problem* of $\mathcal{L}_{DEL}$ is defined as follows:

**Input**: a formula $\varphi \in \mathcal{L}_{DEL}$;

**Output**: yes iff there exists a pointed epistemic model $(\mathcal{M}, w)$ such that $\mathcal{M}, w \models \varphi$.

The satisfiability problem is known to be decidable. Indeed, the standard reduction axioms of DEL [Baltag and Moss, 2004, p. 214] induce a translation $tr : \mathcal{L}_{DEL} \to \mathcal{L}_{EL}$ such that $\varphi \in \mathcal{L}_{DEL}$ is satisfiable iff $tr(\varphi) \in \mathcal{L}_{EL}$ is satisfiable. Since the size of $tr(\varphi)$ is at most exponential in the size of $\varphi$ [Lutz, 2006] and the satisfiability problem of $\mathcal{L}_{EL}$ is PSPACE-complete, the satisfiability problem of $\mathcal{L}_{DEL}$ is in EXPSPACE. This upper bound is nevertheless not optimal: we are going to prove in this section that the satisfiability problem of $\mathcal{L}_{DEL}$ is NEXPTIME-complete.

### 4.1 Upper bound

In this subsection we present a tableau method that does not rely on reduction axioms and we prove that it provides a NEXPTIME procedure deciding the satisfiability problem.

#### 4.1.1 Tableau method

Let $\mathfrak{Lab}$ be a countable set of labels designed to represent worlds of the epistemic model $(\mathcal{M}, w)$. Our tableau method manipulates terms that we call tableau terms and they are of the following kind:

- $(\sigma \quad \mathcal{M}'_1, w'_1; \ldots; \mathcal{M}'_i, w'_i \quad \varphi)$ where $\sigma \in \mathfrak{Lab}$ is a *node* (that represents a world in the initial model) and for all $j \in \{1, \ldots, i\}$, $\mathcal{M}'_j, w'_j$ is an event model. This term means that $\varphi$ is true in the world denoted by $\sigma$ after the execution of the sequence $\mathcal{M}'_1, w'_1, \ldots, \mathcal{M}'_i, w'_i$ and that the sequence is executable in the world denoted by $\sigma$;

- $(\sigma \quad \mathcal{M}'_1, w'_1; \ldots; \mathcal{M}'_i, w'_i \quad \checkmark)$ means that the sequence $\mathcal{M}'_1, w'_1, \ldots, \mathcal{M}'_i, w'_i$ is executable in the world denoted by $\sigma$;

- $(\sigma \quad \mathcal{M}'_1, w'_1; \ldots; \mathcal{M}'_i, w'_i \quad \otimes)$ means that the sequence $\mathcal{M}'_1, w'_1, \ldots, \mathcal{M}'_i, w'_i$ is not executable in the world denoted by $\sigma$;

- $(\sigma R_a \sigma_1)$ means that the world denoted by $\sigma$ is linked by $R_a$ to the world denoted by $\sigma_1$;

- $\bot$ denotes an inconsistency.

A *tableau rule* is represented by a *numerator* $\mathcal{N}$ above a line and a finite list of *denominators* $\mathcal{D}_1, \ldots, \mathcal{D}_k$ below this line, separated by vertical bars:

$$\frac{(\sigma\ \Sigma'\ \varphi \wedge \psi)}{\substack{(\sigma\ \Sigma'\ \varphi)\\(\sigma\ \Sigma'\ \psi)}}\ (\wedge) \qquad \frac{(\sigma\ \Sigma'\ \neg\neg\varphi)}{(\sigma\ \Sigma'\ \varphi)}\ (\neg\neg)$$

$$\frac{(\sigma\ \Sigma'\ \neg(\varphi \wedge \psi))}{(\sigma\ \Sigma'\ \neg\varphi)\mid(\sigma\ \Sigma'\ \neg\psi)}\ (\neg\wedge) \qquad \frac{(\sigma\ \Sigma'\ p)(\sigma\ \Sigma'\ \neg p)}{\bot}\ (\bot)$$

$$\frac{(\sigma\ \Sigma'\ p)}{(\sigma\ \epsilon\ p)}\ (\leftarrow_p) \qquad \frac{(\sigma\ \Sigma'\ \neg p)}{(\sigma\ \epsilon\ \neg p)}\ (\leftarrow_{\neg p})$$

$$\frac{\substack{(\sigma\ \mathcal{M}'_1,w'_1;\ldots;\mathcal{M}'_i,w'_i\ B_a\varphi)\\(\sigma\ R_a\ \sigma_1)}}{(\sigma_1\ \mathcal{M}'_1,u'_1;\ldots;\mathcal{M}'_i,u'_i\ \checkmark)\mid(\sigma_1\ \mathcal{M}'_1,u'_1;\ldots;\mathcal{M}'_i,u'_i\ \otimes)}\ (B_a)$$
$$(\sigma_1\ \mathcal{M}'_1,u'_1;\ldots;\mathcal{M}'_i,u'_i\ \varphi)$$

$$\frac{(\sigma\ \mathcal{M}'_1,w'_1;\ldots;\mathcal{M}'_i,w'_i\ \neg B_a\varphi)}{\substack{(\sigma\ R_a\ \sigma_{\text{new}})\\(\sigma_{\text{new}}\ \mathcal{M}'_1,u'_1;\ldots;\mathcal{M}'_i,u'_i\ \checkmark)\\(\sigma_{\text{new}}\ \mathcal{M}'_1,u'_1;\ldots;\mathcal{M}'_i,u'_i\ \neg\varphi)}}\ (\neg B_a)$$

$$\frac{(\sigma\ \Sigma'\ \neg[\mathcal{M}',w']\varphi)}{\substack{(\sigma\ \Sigma';\mathcal{M}',w'\ \checkmark)\\(\sigma\ \Sigma';\mathcal{M}',w'\ \neg\varphi)}}\ (\neg[\mathcal{M}',w'])$$

$$\frac{(\sigma\ \Sigma'\ [\mathcal{M}',w']\varphi)}{(\sigma\ \Sigma';\mathcal{M}',w'\ \otimes)\mid\substack{(\sigma\ \Sigma';\mathcal{M}',w'\ \checkmark)\\(\sigma\ \Sigma';\mathcal{M}',w'\ \varphi)}}\ ([\mathcal{M}',w'])$$

$$\frac{(\sigma\ \Sigma';\mathcal{M}',w'\ \checkmark)}{\substack{(\sigma\ \Sigma'\ Pre(w'))\\(\sigma\ \Sigma'\ \checkmark)}}\ (\checkmark) \qquad \frac{(\sigma\ \Sigma';\mathcal{M}',w'\ \otimes)}{\substack{(\sigma\ \Sigma'\ \checkmark)\\(\sigma\ \Sigma'\ \neg Pre(w'))}\mid(\sigma\ \Sigma'\ \otimes)}\ (\otimes)$$

$$\frac{(\sigma\ \Sigma'\ \otimes)(\sigma\ \Sigma'\ \checkmark)}{\bot}\ (clash_{\checkmark,\otimes}) \qquad \frac{(\sigma\ \epsilon\ \otimes)}{\bot}\ (\epsilon_\otimes)$$

$$\frac{(\sigma\ \Sigma'\ [\pi\cup\gamma]\varphi)}{\substack{(\sigma\ \Sigma'\ [\pi]\varphi)\\(\sigma\ \Sigma'\ [\gamma]\varphi)}}\ ([\pi\cup\gamma]) \qquad \frac{(\sigma\ \Sigma'\ \neg[\pi\cup\gamma]\varphi)}{(\sigma\ \Sigma'\ \neg[\pi]\varphi)\mid(\sigma\ \Sigma'\ \neg[\gamma]\varphi)}\ (\neg[\pi\cup\gamma])$$

Figure 4: Tableau rules

$$\frac{\mathcal{N}}{\mathcal{D}_1\mid\ldots\mid\mathcal{D}_k}$$

The numerator and the denominators are finite sets of tableau terms.

A *tableau tree* is a finite tree with a set of tableau terms at each node. A rule with numerator $\mathcal{N}$ and denominator $\mathcal{D}$ is *applicable* to a node carrying a set $\Gamma$ if $\Gamma$ contains an instance of $\mathcal{N}$ but not the instance of its denominator $\mathcal{D}$. If no rule is applicable, $\Gamma$ is said to be *saturated*. We call a node $\sigma$ an *end node* if the set of formulas $\Gamma$ it carries is saturated, or if $\bot \in \Gamma$. The tableau tree is extended as follows:

1. Choose a leaf node $n$ carrying $\Gamma$ where $n$ is not an end node, and choose a rule $\rho$ applicable to $n$.

2. (a) If $\rho$ has only one denominator, add the appropriate instantiation to $\Gamma$.

   (b) If $\rho$ has multiple denominators, choose one of them and add to $\Gamma$ the appropriate instantiation of this denominator.

A branch in a tableau tree is a path from the root to an end node. A branch is *closed* if its end node contains $\bot$, otherwise it is *open*. A tableau tree is *closed* if all its branches are closed, otherwise it is *open*. The tableau tree for a formula $\varphi \in \mathcal{L}_{DEL}$ is the tableau tree obtained from the root $\{(\sigma_0\ \epsilon\ \varphi)\}$ when all leafs are end nodes. We write $\vdash \varphi$ when the tableau for $\neg\varphi$ is closed.

The tableau rules of our tableau method are represented in Figure 4. In these rules, $\Sigma'$ is a list of pointed event models $\mathcal{M}'_1,w'_1,\ldots,\mathcal{M}'_i,w'_i$ and $\epsilon$ is the empty list. The tableau method contains the classical Boolean rules $(\wedge),(\neg\neg),(\neg\wedge)$. The rules $(\leftarrow_p)$ and $(\leftarrow_{\neg p})$ handle atomic propositions. The rule $(\bot)$ makes the current execution fail. The rule for $(B_a)$ is applied for all $j \in \{1,\ldots i\}$ and all $u'_j$ such that $w'_j R'_a u'_j$.

Similarly, the rule for $(\neg B_a)$ is applied by choosing non-deterministically for all $j \in \{1,\ldots i\}$ some $u'_j$ such that $w'_j R'_a u'_j$ and creating a new fresh label $\sigma_{\text{new}}$. The rules $(\checkmark)$, $(\otimes)$, $(clash_{\checkmark,\otimes})$ and $(\epsilon_\otimes)$ handle the preconditions. The last two rules $([\pi\cup\gamma])$ and $(\neg[\pi\cup\gamma])$ handle the union operator.

THEOREM 3 (SOUNDNESS AND COMPLETENESS). *Let $\varphi \in \mathcal{L}_{DEL}$. It holds that $\vdash \varphi$ iff $\models \varphi$.*

EXAMPLE 4. *We prove with our tableau method that the formula $\varphi = \neg[\mathcal{M}'_1,w'_1][\mathcal{M}'_2,w'_2]B_2B_1B_2p$ from Example 3 is satisfiable, where $\mathcal{M}'_1,w'_1$ and $\mathcal{M}'_2,w'_2$ are defined in Example 2. In Figure 5, an open branch of the tableau tree for $\varphi$ is represented. The set $\Sigma_{22}$ is saturated: no more tableau rule is applicable. From this branch, we may extract a pointed epistemic model $(\mathcal{M},\sigma_0)$ such that $\mathcal{M},\sigma_0 \models \varphi$.*

### 4.1.2 NEXPTIME-membership

THEOREM 4. *The satisfiability problem of $\mathcal{L}_{DEL}$ is in NEXPTIME.*

PROOF SKETCH. Termination of our tableau method is proved by defining the size of a term $(\sigma\ \Sigma'\ \varphi)$ by $1 + \sum_{(\mathcal{M}',w')\in\Sigma'}(|\mathcal{M}'|+1)+|\varphi|$. The depth of the tableau tree is linear in the size of the input formula, but the number of tableau terms at a node $\sigma$ may be exponential, because of rule $(\neg B_a)$. As a consequence, the tableau tree has at most an exponential number of nodes and constructing non-deterministically such a tree can been done in an exponential amount of time. So, the procedure is in NEXPTIME. $\square$

## 4.2 Lower bound

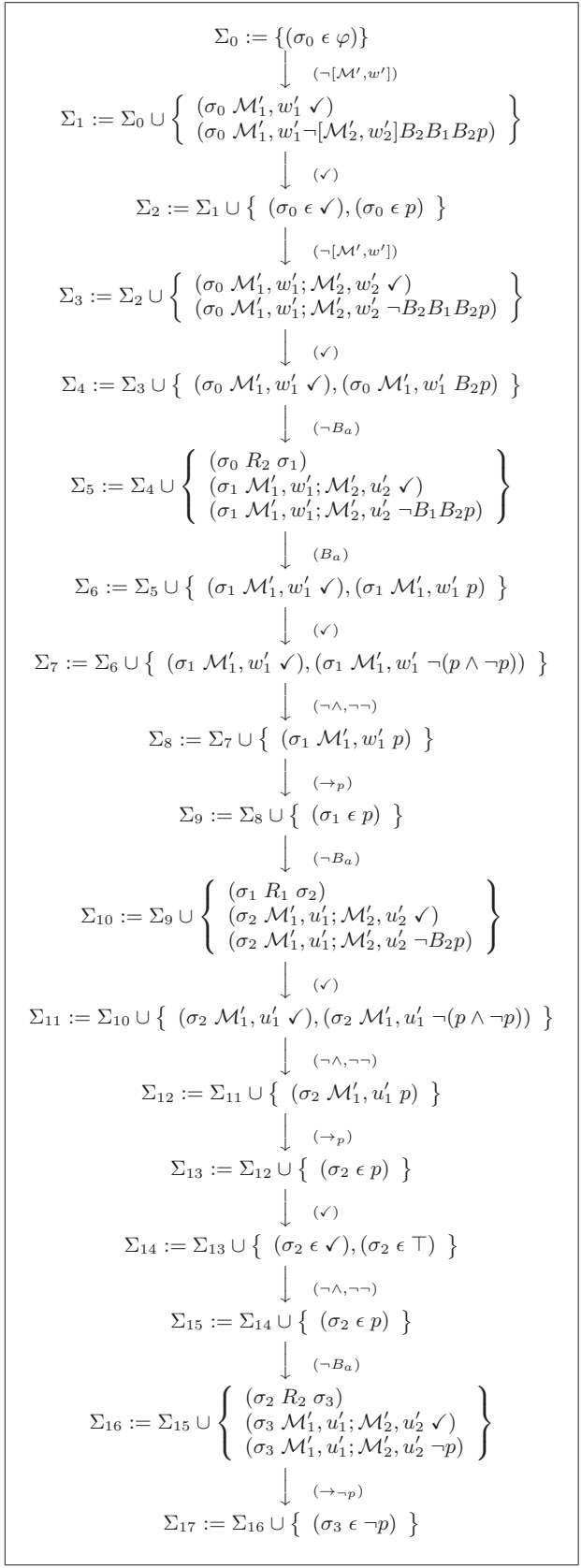

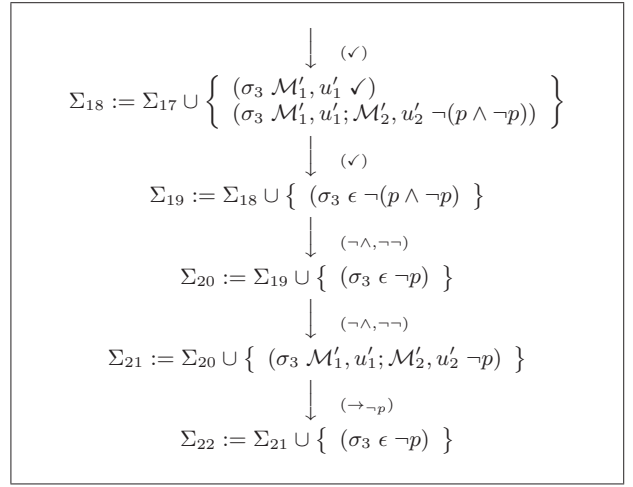

**Figure 5: An open branch of the tableau for $\varphi$**

We prove that the algorithm based on our tableau method of the previous section is optimal in terms of computational complexity. To do so, we prove that the satisfiability problem of $\mathcal{L}_{DEL}$ is NEXPTIME-hard by reducing a NEXPTIME-complete tiling problem to it [Boas, 1997].

Let $C$ be a countable and infinite set of colors. A *tile type* $t$ is a 4-tuple of colors, denoted $t = (left(t), right(t), up(t), down(t)) \in C^4$. We consider the following *tiling problem*:

**Input**: a finite set $T$ of tile types, $t_0 \in T$ and a natural number $k$ written in its binary form.

**Output**: yes iff there exists a function $\tau$ from $\{0, \ldots k\}^2$ to $T$ satisfying the following constraints:

$$\tau(0,0) = t_0; \qquad (1)$$

for all $x \in \{0, \ldots, k\}$ and $y \in \{0, \ldots, k-1\}$:

$$up(\tau(x,y)) = down(\tau(x, y+1)); \qquad (2)$$

for all $x \in \{0, \ldots, k-1\}$ and $y \in \{0, \ldots, k\}$:

$$right(\tau(x,y)) = left(\tau(x+1, y)). \qquad (3)$$

In other words, the problem is to decide whether we can tile a $(k+1) \times (k+1)$ grid with the tile types of $T$, $t_0$ being placed onto $(0,0)$.

THEOREM 5. *The satisfiability problem of $\mathcal{L}_{DEL}$ is NEXPTIME-hard.*

PROOF. Without loss of generality, we assume that $k = 2^n$. Let us consider an instance of the NEXPTIME-hard tiling problem described above. Our goal is to provide a polynomial translation from this instance to an instance of the satisfiability problem of $\mathcal{L}_{DEL}$.

The idea is to embed *two identical* $k \times k$-tilings into a single tree. Each leaf of the tree represents both a position $(x_1, y_1)$ in the first tiling and a position $(x_2, y_2)$ in the second tiling. We need to encode *two identical* tilings because, in order to check constraints 2 and 3, we will need to refer to the tile located to the right or to the left of a given position in a tiling, and also to refer to the tile located above or below it. This is hardly possible if we encode a single tiling at the leafs of a tree, because we would need to 'backtrack' in the tree to access these other positions.

We start by showing how to encode two identical tilings at the leafs of a tree. Then we will show how to express the three constraints 1, 2 and 3 in the definition of a tiling.

1. The coordinates $(x_1, y_1)$ and $(x_2, y_2)$ of the two tilings are represented by the valuations of atomic propositions $p_0, \ldots, p_{4n-1}$. More precisely, the set $X_1 = \{p_0, \ldots, p_{n-1}\}$ contains the atomic propositions encoding the binary representation of the integer $x_1$, $Y_1 = \{p_n, \ldots, p_{2n-1}\}$ contains the atomic propositions encoding the binary representation of the integer $y_1$, $X_2 = \{p_{2n}, \ldots, p_{3n-1}\}$ contains the atomic propositions encoding the binary representation of the integer $x_2$, and $Y_2 = \{p_{3n}, \ldots, p_{4n-1}\}$ contains the atomic propositions encoding the binary representation of the integer $y_2$. For instance, for $n = 4$, the coordinates $(x_1, y_1) = (4, 3)$ and $(x_2, y_2) = (11, 2)$ are represented at a leaf of the tree by the following valuation. We recall that in binary notation, 4 is represented by $\overline{100}$, 3 is represented by $\overline{11}$, 12 is represented by $\overline{1100}$ and 2 is represented by $\overline{10}$.

$$\underbrace{\neg p_0, p_1, \neg p_2, \neg p_3}_{4} \underbrace{\neg p_4, \neg p_5, p_6, p_7}_{3}$$

$$\underbrace{p_8, p_9, \neg p_{10}, \neg p_{11}}_{12} \underbrace{\neg p_{12}, \neg p_{13}, p_{14}, \neg p_{15}}_{2}$$

We then encode the existence of all valuations over $X_1 \cup Y_1 \cup X_2 \cup Y_2$ with the following formula:

$$\bigwedge_{l < 4n} B_a^l \left( \langle B_a \rangle p_l \wedge \langle B_a \rangle \neg p_l \wedge \bigwedge_{i < l} ((p_i \to B_a p_i) \wedge (\neg p_i \to B_a \neg p_i)) \right). \quad (4)$$

Formula 4 is true at a pointed epistemic model iff this pointed epistemic model is bisimilar up to modal depth $4n$ to a binary tree of depth $4n$ whose leafs contain all the possible valuations associated to $p_0, \ldots, p_{4n-1}$.

In order to check Constraints 2 and 3 in the definition of a tiling, we will need to refer to the tile located to the right or to the left of a given position in a tiling, and also to refer to the tile located above or below it. The following formulas encode the fact that any pair of coordinates $(x_1, x_2)$ and $(y_1, y_2)$ of the two tilings satisfy the properties $x_1 = x_2$, $x_1 = x_2 + 1$, $y_1 = y_2$ and $y_1 = y_2 + 1$ respectively:

$$(x_1 = x_2) \triangleq \bigwedge_{i < n} (p_i \leftrightarrow p_{i+2n}) \quad (5)$$

$$(y_1 = y_2) \triangleq \bigwedge_{n \leq i < 2n} (p_i \leftrightarrow p_{i+2n}) \quad (6)$$

$$(x_1 = x_2 + 1) \triangleq \bigvee_{i < n} \left( \bigwedge_{j < i} (p_{j+2n} \leftrightarrow p_j) \wedge \neg p_{i+2n} \wedge p_i \right.$$
$$\left. \wedge \bigwedge_{i < j < n} (p_{j+2n} \wedge \neg p_j) \right) \quad (7)$$

$$(y_1 = y_2 + 1) \triangleq \bigvee_{n \leq i < 2n} \left( \bigwedge_{n \leq j < i} (p_{j+2n} \leftrightarrow p_j) \wedge \neg p_{i+2n} \wedge p_i \right.$$
$$\left. \wedge \bigwedge_{i < j < 2n} (p_{j+2n} \wedge \neg p_j) \right) \quad (8)$$

The tile types of the first tiling are represented by atomic propositions $1_t$ and the tile types of the second tiling are represented by atomic propositions $2_{t'}$, where $t$ and $t'$ range over $T$. They hold at a leaf of the tree whose coordinates correspond to $(x_1, y_1)$ and $(x_2, y_2)$ when the tile type of the first tiling at coordinate $(x_1, y_1)$ is $t$ and the tile type of the second tiling at coordinate $(x_2, y_2)$ is $t'$.

Formulas 9 and 10 below encode the fact that, *at each leaf of the tree*, there is exactly one tile type for the first tiling and exactly one tile type for the second tiling. Formula 11 below encodes the fact that when these two pairs of coordinates coincide, that is when $x_1 = x_2$ and $y_1 = y_2$, then the tile type of the first tiling and the tile type of the second tiling are identical.

$$B_a^{4n} \left( \bigvee_{t \in T} 1_t \wedge \bigvee_{t \in T} 2_t \right) \quad (9)$$

$$B_a^{4n} \bigwedge \{(1_t \to \neg 1_{t'}) \wedge (2_t \to \neg 2_{t'}) \mid t, t' \in T, t \neq t'\} \quad (10)$$

$$B_a^{4n} \left( (x_1 = x_2) \wedge (y_1 = y_2) \to \bigwedge_{t \in T} (1_t \leftrightarrow 2_t) \right) \quad (11)$$

However, it may be the case that in the tree, two different leafs with the *same* valuation have different tile types. Therefore, we also have to constrain the tree so that the leafs denoting the same position in the first tiling (resp. second tiling) contain the same tile type for the first tiling (resp. second tiling). This is expressed by the following two formulas:

$$[\mathcal{M}'_{p_0} \cup \mathcal{M}'_{\neg p_0}] \ldots [\mathcal{M}'_{p_{2n-1}} \cup \mathcal{M}'_{\neg p_{2n-1}}] \bigvee_{t \in T} B_a^{4n} 1_t \quad (12)$$

$$[\mathcal{M}'_{p_{2n}} \cup \mathcal{M}'_{\neg p_{2n}}] \ldots [\mathcal{M}'_{p_{4n-1}} \cup \mathcal{M}'_{\neg p_{4n-1}}] \bigvee_{t \in T} B_a^{4n} 2_t \quad (13)$$

where for a given a literal $\ell$ ($p$ or $\neg p$), the pointed event model $\mathcal{M}'_\ell = (W', R', Pre, w'_0)$ is defined as follows: $W' = \{w'_i \mid i \in \{0, \ldots, 4n\}\}$; $R'_a = \{(w'_i, w'_{i+1}) \mid i \in \{0, \ldots, 4n-1\}\}$; and $Pre(w'_i) = \top$ for all $i < 4n$ and $Pre(w'_{4n}) = \ell$.

In formula 12, the sequence of pointed event models $[\mathcal{M}'_{p_0} \cup \mathcal{M}'_{\neg p_0}] \ldots [\mathcal{M}'_{p_{2n-1}} \cup \mathcal{M}'_{\neg p_{2n-1}}]$ non-deterministically picks a valuation $v$ over $X_1 \cup Y_1$ and selects the branches of the tree whose leafs satisfy this valuation. Then, the formula $\bigvee_{t \in T} B_a^{4n} 1_t$ checks that these leafs, which denote the same position in the first tiling, are of the same tile type $t$. Likewise with formula 13 for the second tiling.

So, with formulas 9, 10, 11, 12 and 13, we have encoded in the tree two identical tilings in a single tree. Importantly, note that the tree is defined so that each leaf refers to two coordinates of the tiling, which can possibly be identical or consecutive. It is this feature which will allow us to express that constraints 2 and 3 of the definition of a tiling hold.

2. Constraints 1, 2 and 3 of the definition of a tiling are expressed respectively by the following formulas:

$$B_a^{4n} \left( \left( \bigwedge_{i < 4n} \neg p_i \right) \to t_0 \right) \quad (14)$$

$$B_a^{4n} \left( (x_1 = x_2) \wedge (y_1 = y_2 + 1) \right. \quad (15)$$
$$\left. \to \bigwedge_{t \in T} \left\{ 1_t \to \bigvee \{2_{t'} \mid t' \in T, down(t') = up(t)\} \right\} \right)$$

$$B_a^{4n}\bigg((x_1 = x_2 + 1) \wedge (y_1 = y_2) \qquad (16)$$
$$\rightarrow \bigwedge_{t \in T} \Big\{1_t \rightarrow \bigvee \{2_{t'} \mid t' \in T, left(t') = right(t)\}\Big\}\bigg)$$

As we said at the beginning of the proof, these two constraints motivate the need to encode *two* tilings: for a given position in a tiling, we need to refer to the tile located to the right or to the left of it, and to refer to the tile located above or below it. This would not be possible with our epistemic language if the tiling was encoded by a single tree.

One can then check that there exists a tiling for the instance of the tiling problem iff the formula $\varphi$, which is the conjunction of fomulas 4, 9, 10, 11, 12, 13, 14, 15, and 16 is satisfiable in $\mathcal{L}_{DEL}$.

3. Finally, we show that the reduction is polynomial in the size of the instance of the tiling problem. The formula of Equation 4 is of size $O(n^2)$. The formulas of Equations 12, 13 are of size $O(n^2 + |T| \times n)$. The other formulas are clearly of size polynomial in the size of the input, so the result follows. Importantly, note that if we decided to rewrite the formulas 12 and 13 without using the union operator $\cup$, then the corresponding formula would be exponential in the size of the input. So, the use of the union operator is really crucial in order to have a polynomial reduction from the tiling problem to our satisfiability problem. □

# 5. RELATED WORK

## 5.1 Theory

There exists a terminating tableau method solving the satisfiability problem of $\mathcal{L}_{DEL}$ [Hansen, 2010]. This method writes subformulas by applying the reduction axioms [Baltag and Moss, 2004, p. 214]. It is therefore mainly a variant of the tableau method of classical multi-modal logic $K_n$. Even if we know that $tr$ blows up exponentially the size of the input formula, the computational complexity of this tableau method is not studied. In this section, we review the existing results about computational complexity of DEL.

### 5.1.1 Public Announcement Logic (PAL)

Public Announcement Logic (PAL) [Plaza, 1989] is an extension of epistemic logic with a dynamic operator $[\psi!]\varphi$ whose truth conditions are defined as follows:

$$\mathcal{M}, w \models [\psi!]\varphi \quad \text{iff} \quad \mathcal{M}, w \models \psi \text{ implies } \mathcal{M}_\psi, w \models \varphi$$

where $\mathcal{M}_\psi$ is the restriction of $\mathcal{M}$ to the worlds which satisfy $\psi$. PAL is a fragment of DEL: the language of PAL is $\mathcal{L}_{DEL}$ restricted to event models consisting of a single possible event with reflexive arrows for all agents. There is a gap between PAL and DEL in terms of computational complexity, both for the model checking problem and the satisfiability problem. Indeed, the model checking of PAL is in P (also with common belief) [van Benthem and Kooi, 2004] and the satisfiability problem for PAL is PSPACE-complete [Lutz, 2006]. Despite the fact that there exist reduction axioms for PAL, it is difficult to implement a direct translation using reduction axioms. In fact, there are properties that can be expressed exponentially more succinctly in PAL than in epistemic logic [French et al., 2011]. Note that there exist PSPACE tableau methods for solving the satisfiability problem in PAL [de Boer, 2007, Balbiani et al., 2010].

### 5.1.2 DEL-sequents

DEL-sequents [Aucher, 2011] are triples of the form $\varphi, \varphi' \models \varphi''$ where $\varphi, \varphi'' \in \mathcal{L}_{EL}$ and $\varphi'$ is a formula of a language for event models. A DEL-sequent $\varphi, \varphi' \models \varphi''$ holds when for all pointed epistemic model $(\mathcal{M}, w)$ such that $\mathcal{M}, w \models \varphi$, for all pointed event model $(\mathcal{M}', w')$ such that $\mathcal{M}', w' \models \varphi'$, if $(\mathcal{M}', w')$ is executable in $(\mathcal{M}, w)$, then $\mathcal{M} \otimes \mathcal{M}', (w, w') \models \varphi''$. The problem of determining whether a DEL-sequent holds is NEXPTIME-complete and there exists a tableau method for it. DEL-sequents have been generalized to sequences of the form $\varphi_0, \varphi'_1, \varphi_1, \ldots, \varphi'_n, \varphi_n \models^1_i \psi$ and $\varphi_0, \varphi'_1, \varphi_1, \ldots, \varphi'_n, \varphi_n \models^2_i \psi'$. The corresponding satisfiability problem is also NEXPTIME-complete [Aucher et al., 2012].

### 5.1.3 The sequence and 'star' iteration operators

The sequence and 'star' iteration operators are constructions enabling to build complex programs as in Propositional Dynamic Logic (PDL [Harel et al., 2000]). The truth conditions are defined as follows:

$$\mathcal{M}, w \models [\pi; \gamma]\varphi \quad \text{iff} \quad \mathcal{M}, w \models [\pi][\gamma]\varphi$$
$$\mathcal{M}, w \models [\pi^*]\varphi \quad \text{iff} \quad \text{there is a } \textit{finite} \text{ sequence } \pi; \ldots; \pi$$
$$\text{such that } \mathcal{M}, w \models [\pi; \ldots; \pi]\varphi$$

We do not know about the computational complexity of the model-checking problem when the operator $[\pi^*]\varphi$ is added to the language. In fact, we do not even know whether it is decidable. The computational complexity of the satisfiability problem remains the same when the sequential composition operator is added. However, adding a 'star' operator makes the satisfiability problem undecidable. This result is not really surprising, it is a direct corollary of the result of [Miller and Moss, 2005] stating that Public Announcement Logic with the 'star' operator is already undecidable.

### 5.1.4 The common belief operator

We may extend the language with the common belief operator $C_G\varphi$, where $G \subseteq AGT$. The truth conditions are defined as follows:

$$\mathcal{M}, w \models C_G\varphi \quad \text{iff} \quad \text{for all } v \in \bigg(\bigcup_{a \in G} R_a\bigg)^+ (w), \mathcal{M}, v \models \varphi$$

Intuitively, $C_G\varphi$ is an abbreviation of an infinite conjunction [Fagin et al., 1995]: $C_G\varphi = E_G^1\varphi \wedge E_G^2\varphi \wedge E_G^3\varphi \wedge \ldots$, where $E_G^k\varphi$ is defined inductively as follows: $E_G^1\varphi = \bigwedge_{a \in G} B_a\varphi$ and $E_G^{k+1}\varphi = E_G^1 E_G^k \varphi$.

We do not know about the computational complexity of the satisfiability problem when the common belief operator is added to the language $\mathcal{L}_{DEL}$. However, we know that it is decidable and that the language with common belief operator is more expressive than the *epistemic* language $\mathcal{L}_{EL}$ with common belief [Baltag et al., 1998, Baltag et al., 1999].

## 5.2 Implementation

There exist two implementations of our decision problems:
1. The model-checker DEMO [van Eijck, 2007], standing for Dynamic Epistemic MOdeling tool, can evaluate formulas of $\mathcal{L}_{DEL}$ in epistemic models, display graphically epistemic models, event models and updates of epistemic models by event models, translate formulas of $\mathcal{L}_{DEL}$ to formulas of PDL. DEMO is written in Haskel and has been applied in [van Ditmarsch et al., 2005] and [van Ditmarsch et al., 2006]. Also, it has been used to investigate the pros and cons of modeling some well-known problems of computer security within the DEL framework [van Eijck and Orzan, 2007].

2. The program `Aximo` [Richards and Sadrzadeh, 2009], written in C++, implements an algorithm for proving properties of interactive multi-agent scenarios encoded in epistemic systems. Epistemic systems provide an algebraic semantics to DEL and were developed together with a sound and complete sequent calculus [Baltag et al., 2007].

# 6. CONCLUDING REMARKS

Our work contributes to the proof theory and the study of the computational complexity of DEL, which has been rather neglected so far. Although our results show that our decision problems are not tractable, it turns out that the DEMO implementation does not fare worse and often even better in terms of time of execution than other model-checkers modeling the same problems, without resorting to the DEL methodology [van Ditmarsch et al., 2006].

We still need to investigate whether or not the computational complexity remains the same when we consider other epistemic logics as the basis of DEL, such as S5. Moreover, our results rely on the fact that we use the union operator in the language, an open problem is to obtain similar results without this operator. Finally, we plan to implement our tableau method in LotrecScheme [Schwarzentruber, 2011].

**Acknowledgment.** We thank the reviewers for their comments and Sophie Pinchinat for discussions.